\documentclass[11pt,letterpaper]{article}
 \pdfoutput=1

\usepackage{jcappub}

\usepackage{caption}
\usepackage{floatrow}
\usepackage{journals}

\usepackage{subfig}

\usepackage{bm}

\def\be{\begin{eqnarray}}
\def\ee{\end{eqnarray}}

\def\k{\bm{k}}

\def\x{\bm{x}}

\def\q{\bm{q}}
\def\s{\bm{s}}

\def\d{\delta_D}

\def\mathdsOne{\mathrm{I}}

\newsavebox{\tempbox}

\title{N-point Statistics of Large-Scale Structure in the Zel'dovich Approximation}
\author{Svetlin Tassev$^{a}$}

\affiliation{ \sl $^{a}$ Department of Astrophysical Sciences, Princeton University, 4 Ivy Lane, Princeton, \\NJ 08544, USA
}

\abstract{Motivated by the results presented in a companion paper, here we give a simple analytical expression for the matter $n$-point functions in the Zel'dovich approximation (ZA) both in real and in redshift space (including the angular case). We present numerical results for the 2-dimensional redshift-space correlation function, as well as for the  equilateral configuration for the real-space 3-point function. We compare those to the tree-level results. Our analysis is easily extendable to include Lagrangian bias, as well as higher-order perturbative corrections to the ZA. The results should be especially useful for modelling probes of large-scale structure in the linear regime, such as the Baryon Acoustic Oscillations. We make the numerical code used in this paper freely available.
}

\usepackage{graphicx}
\begin{document}
\maketitle

\section{Introduction}

Probes of Large-Scale Structure (LSS) have the potential to give powerful constraints on dark energy and dark matter (e.g. \cite{2013PhR...530...87W}). One such measurement is the accurate determination of the Baryon Acoustic Oscillations (BAO) scale. However, future surveys such as WFIRST, Euclid and LSST will provide precise measurements of the acoustic peak in the matter 2-point correlation function at a level which our current theoretical understanding of LSS does not yet completely match. 

Using N-body simulations to model the acoustic peak, and especially the  uncertainties behind those measurements, requires running hundreds and even thousands of simulations \cite{2013MNRAS.428.1036M}.  Therefore, theoretical models of the BAO signal and its  covariance can still play an essential role for extracting cosmological data from  observations. 

In a companion paper \cite{2013arXiv1311.4884T}, we demonstrated that models of LSS can be improved beyond the level of accuracy required for future surveys, as long as one avoids Fourier space and builds a perturbative model around the Zel'dovich Approximation (ZA) \cite{zeldovich}. However, this result would be of little value if even the simplest of such models -- the ZA itself -- cannot provide analytical\footnote{Meaning, predictions which do not require numerical averaging over realizations.} predictions beyond the 2-pt correlation function, such as its covariance. Indeed, until now the ZA has had a very serious drawback -- analytical solutions existed only for the 2-point correlation function (e.g. \cite{2013MNRAS.429.1674C}). Thus, the ZA has mostly been regarded as a cheap way of performing crude N-body simulations (e.g. \cite{2013MNRAS.428.1036M}), and rarely as an analytical power horse. 



In an attempt to change that, in this paper we give a simple  analytical expression for the  matter $n$-point functions in the ZA both in real and redshift space.\footnote{Following the discussion we present in \cite{2013arXiv1311.4884T}, we avoid Fourier space at all cost.}
Thus, we show that  for  models perturbing around the ZA solution, the lowest order (the ZA itself) gives well defined analytical predictions for the statistics of  Cold Dark Matter (CDM). 

We implement  our result numerically in  an open-source code, called  Zel'dovich Calculator (ZelCa\footnote{In case the reader is curious, the word means \textit{cabbage} in Bulgarian. The code is freely available at  the following URL: \url{https://bitbucket.org/tassev/zelca/}}). The code is capable of calculating  the real and redshift space 2-pt functions, as well as the real-space 3-pt function in the ZA.
We include results for those quantities in this paper.

In Section~\ref{sec:za} we give a quick overview of the Zel'dovich approximation and introduce our notation. We use Section~\ref{sec:2pt} to warm up with the standard calculation for the 2-pt function, and then in Section~\ref{sec:npt} we derive our expression for the matter $n$-point functions in real and redshift space. In Section~\ref{sec:numdisc} we describe the numerical implementation of our results in ZelCa and discuss current numerical limitations. In Section~\ref{sec:nums} we present our numerical results. We summarize in Section~\ref{sec:summary}, where we discuss how our results can be extended to include Lagrangian bias as well as higher-order corrections (including corrections arising when the theory is renormalized, see \cite{2013arXiv1311.4884T}).


\section{ The Zel'dovich approximation}\label{sec:za}

Let us introduce some notation by doing a quick overview of the ZA. A CDM particle with an initial (Lagrangian) position $\q$ ends up at a final (Eulerian) position $\x$ after time $t$ according to:
\be
\x(\q,t)=\q+\s(\q,t)
\ee
with $\s$ being the so-called displacement field. The above equation is valid in the  general case. The ZA boils down to using the linear result for the displacement field given by $\s(\q,t)=D(t)\s_0(q)$, where $D$ is the linear growth factor. The displacement field in the ZA is a Gaussian random variable with zero mean. Its variance is given by
\be\label{psi}
\psi_{ij}(\q_a-\q_b,t)\equiv\langle s_i(\q^a,t)s_j(\q^b,t)\rangle=\int \frac{d^3k}{(2\pi)^3} e^{i\k\cdot(\q^a-\q^b)}\frac{k_ik_j}{k^4}P_L(k,t) 
\ee
where $P_L(k,t)$ is the linear power spectrum at time $t$. It is given by $P_L(k,t)=D(t)^2 P_{L}(k,t_0)$, where $t_0$ is defined by $D(t_0)=1$. After some algebra, the above expression can be written as (we drop the time arguments for brevity)
\be\label{psiEq}
\psi_{ij}(\q)=\chi(q) \delta_{ij}-3\gamma(q)\hat q_i \hat q_j \ ,
\ee
where
\be
\chi(q)\equiv \frac{4 \pi}{3}\int \frac{dk}{(2\pi)^3}  P_L(k,t) \left(j_0(q k)+j_2(qk)\right) \ , \ \hbox{and} \ \ 
\gamma(q)\equiv \frac{4 \pi}{3}\int \frac{dk}{(2\pi)^3} P_L(k,t) j_2(qk)
\ee
Note that 
\be\label{psi0}
\psi_{ij}(\q=0)=\delta_{ij}\sigma^2 \ ,
\ee
 where
\be
\sigma^2=\frac{4\pi}{3}\int \frac{dk}{(2\pi)^3} P_L(k)
\ee

\subsection{Real space}

The CDM overdensity field, $\delta$,  is given by
\be\label{beginning}
1+\delta(x,t)=\int d^3q \d(\x-\q-\s(q,t))
\ee
which can also be rewritten as
\be\label{density}
1+\delta(x,t)=\int \frac{d^3q d^3k}{(2\pi)^3}\, e^{-i\k\cdot(\x-\q-\s(q,t))}
\ee
From now on we drop the arguments of the functions above for brevity.

\subsection{Redshift space}

In redshift space, the density in the ZA is given by 
\be
1+\delta_{(r)}(x,t)=\int d^3q \d(\x-\q-\s_{(r)}(q,t))\ ,\ 
\ \hbox {with}\nonumber\\ 
\s_{(r)}\equiv \s(\q,t)+f\s_{||}(\q,t)
\ee
where $f$ is the dimensionless linear growth rate, $f\equiv d\ln D/d\ln a$, $a$ being the scale factor. Subscript $(r)$ denotes redshift space. Above we used $\s_{||}\equiv (\s\cdot\hat r)\hat r$, where $\hat r$ is the direction along the line-of-sight. Thus, $\s_{(r)}$ implicitly depends on the $\x$ for which we evaluate $\delta_{(r)}$.

In that case, we can redo the algebra of the previous section to find that nothing changes, except $\psi$ now has to be replaced with its redshift-space counterpart, $\psi_{(r)}$, which can be easily seen to be equal to:
\be\label{psir}
\psi_{(r)}(\q,\x^a,\x^b)&=&\langle \s_{(r)}(\q_0)\s_{(r)}(\q_0+\q) \rangle = L^{a}\psi(\bm{q}) L^{b}\ \ , \ \ \hbox{where}\nonumber\\
L^{a}_{ij}&\equiv&(\delta_{ij}+f\hat r^{a}_i\hat r^{a}_j)\ \ ,
\ee
where the superscripts keep track of the $\x^{a}$  towards which the line-of-sight vector $\hat r^a$ is pointing.
Note that the above expression is valid in both the plane-parallel (for which $\hat r$ is constant, and the superscripts can be dropped as well as the $\x$ dependence) and in the angular case of redshift-space distortions. 

\section{The 2-point function in real and redshift space}\label{sec:2pt}
Let us warm up with deriving the 2-pt function, $\xi(r)$, which is well-known in the literature (e.g. \cite{2013MNRAS.429.1674C}).  We will go step by step, because deriving the $n$-point functions will follow the same logic. 

Let us first start with real space. Using (\ref{density}), the 2pt function in the ZA can be written as:
\be\label{xiInit}
\xi(|\x^b-\x^a|)&=&\langle \delta (\x^a)\delta (\x^b)\rangle-\langle \delta (\x^a)\rangle\langle \delta (\x^b)\rangle=\\
&=&\int d^3k^a d^3k^b e^{-i(\k^a\cdot\x^a+\k^b\cdot\x^b)}\int\frac{d^3 q^a d^3q^b}{(2\pi)^6}
e^{i(\k^a\cdot\q^a+\k^b\cdot\q^b)}\times\nonumber\\
&\times&\left[e^{-\frac{1}{2}
\bigg\langle \left(\k^a\cdot\s^a+\k^b\cdot\s^b\right)^2
\bigg\rangle_c}-e^{-\frac{1}{2}
\bigg\langle \left(\k^a\cdot\s^a\right)^2+ \left(\k^b\cdot\s^b\right)^2
\bigg\rangle_c}
\right]
\nonumber
\ee
where $\s^a\equiv \s(\q^a)$ (and in general, vector labels in this paper will appear as superscripts, whereas vector indices  -- as subscripts). In writing the last line of the above equation, we used that for any Gaussian random variable, $A$, with zero mean, the cumulant expansion gives us:
\be
\langle e^A\rangle=e^{\frac{1}{2}\langle A^2\rangle_c}
\ee

Plugging (\ref{psi},\ref{psi0}) in (\ref{xi}) we find
\be
\xi(|\x^b-\x^a|)&=&\int d^3k^a d^3k^b e^{-i(\k^a\cdot\x^a+\k^b\cdot\x^b)}\int\frac{d^3 q^a d^3q^b}{(2\pi)^6}
e^{i(\k^a\cdot\q^a+\k^b\cdot\q^b)}\times\nonumber\\
&\times&
e^{-\frac{1}{2}\sigma^2 \left((k^{a})^2+(k^{b})^2\right)}
\left[e^{-k^a_ik^b_j\psi_{ij}(\q_b-\q_a)}-1
\right]
\ee
where repeated subscripts are summed over.
Now let's change variables to $(\q^a,\q^b)\to (\q^a,\q^{ab}\equiv\q^a-\q^b)$. We can then do the integral in $\q^a$ which gives a delta function, removing one of the $\k$ integrals. We then find
\be
\xi(|\x^b-\x^a|)&=&\int d^3k e^{-i\k\cdot(\x^a-\x^b)}\int\frac{d^3q^{ab}}{(2\pi)^3}
e^{i\k\cdot\q^{ab}}
e^{-\sigma^2 k^2}
\left[e^{k_ik_j\psi_{ij}(\q^{ab})}-1
\right]
\ee
Now we perform the Gaussian integral in $\k$ to obtain
\be\label{xi}
1+\xi(|\x^{ab}|)=\int\frac{d^3q^{ab}}{(2\pi)^{3/2}}\frac{1}{\sqrt{\det\left[N_{ij}(\q^{ab})\right]}}e^{-\frac{1}{2}(x^{ab}_i-q^{ab}_i)\left[N(\q^{ab})^{-1}\right]_{ij}(x^{ab}_j-q^{ab}_j)}
\ee
where  $\x^{ab}\equiv \x^a-\x^b$ and
\be\label{N}
\frac{1}{2}N_{ij}(\q)\equiv \sigma^2\delta_{ij}-\psi_{ij}(\q)
\ee
with $\psi_{ij}$ given by (\ref{psiEq}).

Fixing the $\x^{ab}$ to be our $\hat z$ axis, then the azimuthal angular integral of $d^3q^{ab}$ trivially gives $2\pi$, and we are left with a 2-dimensional integral to evaluate numerically\footnote{Evaluating the polar angle integral analytically is also possible, but then the expression can no longer be easily compared to the general $n$-point function result.}:  in $|q^{ab}|$ and in the polar angle, the one between $\q^{ab}$ and $\x^{ab}$. 

In redshift space, in the angular case $\xi(\x^a,\x^b)$ is a function of both $\x^a$ and $\x^b$ and not only their difference as translation invariance is broken; while in the plane-parallel case one has $\xi(\x^{ab})$ depending on the direction of $\x^{ab}$ as isotropy is broken. Keeping in mind those two things,  one can see that the above calculation is followed through transparently, except that now one has to use the redshift space $N$ in (\ref{xi}) defined through:
\be\label{Nr}
N_{(r),ij}(\q,\x^a,\x^b)\equiv \psi_{(r),ij}(\bm{0},\x^a,\x^a)+\psi_{(r),ij}(\bm{0},\x^b,\x^b)-\psi_{(r),ij}(\q,\x^a,\x^b)-\psi_{(r),ij}(\q,\x^b,\x^a)\nonumber\\
\ee
The above equation is again valid both for the plane-parallel as well as for the angular case. Note that $\x^a$ appears alone (i.e. not as the difference $\x^{ab}$) only in $N_{(r)}$ through $\psi_{(r)}$, which in turn depends on $\hat r^a$. Moreover, the dimensionality of the integral remains unchanged in redshift space.

\section{General $n$-point functions in real  and redshift space}\label{sec:npt}

In this section we will give a general expression for the $n$-point functions of $\delta$. We start by noting that the $n$-point function of $\delta$ equals the $n$-point function of $(1+\delta)$, since a shift in the mean value of a random variable does not change its cumulants. Any $n$-point function  can therefore be expressed through a linear combination of the moments (which include connected and disconnected pieces)  $\langle (1+\delta)^m\rangle$, with $m\leq n$. Thus, if we know $\langle (1+\delta)^m\rangle$, we can easily obtain the desired $n$-point function of $\delta$.

So, in this section we will calculate the moments $G_n\equiv\langle (1+\delta)^n\rangle$, or more explicitly:
\be
G_{n}(\x^{1},\x^{2},\x^{3}...,\x^{n})\equiv \langle (1+\delta(\x^1))(1+\delta(\x^2))(1+\delta(\x^3))...(1+\delta(\x^n))\rangle\ ,
\ee
where numerical superscripts allow us to distinguish the different $\x$'s. Note that even though we could set $\x^1=0$ without loss of generality in real space, we would like to keep the discussion applicable in the angular case in redshift space. So, we keep $\x^1$  in $G_n$. However, note that even in the angular case in redshift space, we still can use translation invariance in Lagrangian space -- a fact which will simplify greatly our final result.

To make the notation more compact, let us construct the following column vectors of length $3n$ by stacking $\x$, $\q$, etc.:
\be\label{Vecs}
\tilde{\bm{X}}^T&\equiv& \left((\x^1)^T,(\x^2)^T,(\x^3)^T,...(\x^n)^T\right)\nonumber\\
\tilde{\bm{Q}}^T&\equiv& \left((\q^1)^T,(\q^2)^T,(\q^3)^T,...(\q^n)^T\right)\nonumber\\
\bm{S}^T&\equiv& \left((\s^1)^T,(\s^2)^T,(\s^3)^T,...(\s^n)^T\right)\nonumber \\
\tilde{\bm{K}}^T&\equiv& \left((\k^1)^T,(\k^2)^T,(\k^3)^T,...(\k^n)^T\right)\ ,
\ee
where as before $\s^a\equiv \s(\q^a)$, and a superscript $T$ stands for the matrix transpose. Note that $\s$ here  stands  either  for the real-space $\s$ or for  its redshift-space counterpart, $\s_{(r)}$. The role of the tildes  will become apparent below. As an example, if $n=2$, then ${\bm{S}}^T=(s_1^1,s_2^1,s_3^1,s_1^2,s_2^2,s_3^2)$, where again a superscript denotes a vector label, while subscript -- the vector index. Therefore, as an example, the fourth element of $\bm S$ is given by $S_4=s_1^2$.

With the above notation, we can automatically write down $G_n$ in the same way we calculated the 2-pt function:
\be\label{Gtemp}
G_n=\int \frac{d^{3n}\tilde Qd^{3n}\tilde K}{(2\pi)^{3n}} e^{-i \tilde{\bm{K}}\cdot (\tilde{\bm{X}}-\tilde{\bm{Q}})} e^{-\frac{1}{2}\tilde K_i \langle S_i S_j \rangle \tilde K_j}
\ee
where the subscripts of $\bm{\tilde K}$, $\bm S$, etc. run over both the subscripts and superscripts of $\k$, $\s$, etc. as per their definition (\ref{Vecs}). 

Let us define the displacement covariance:
\be\label{Mold}
M_{ij}(\tilde{\bm{Q}})\equiv \langle S_i S_j \rangle\ ,
\ee
Explicitly in real space it is given by the following block form:
\be\label{M}
M=\begin{pmatrix}
  \sigma^2 & \psi^{21} & \psi^{31} & \cdots & \psi^{n1} \\
  \psi^{21} & \sigma^2 & \psi^{32} & \cdots & \psi^{n2} \\
  \psi^{31} & \psi^{32} & \sigma^2  & \cdots & \psi^{n3} \\
  \vdots  & \vdots & \vdots & \ddots & \vdots  \\
  \psi^{n1} & \psi^{n2} & \psi^{n3} &  \cdots & \sigma^2 
 \end{pmatrix}
\ee
where each block is a 3-by-3 matrix. We defined $\q^{ab}\equiv\q^{a}-\q^{b}$. Implicitly each $\sigma^2$ is multiplied by $\mathdsOne_{3\times 3}$, the identity matrix.
The 3-by-3 matrices $\psi^{ab}\equiv \psi(\q^{ab})$ are given by (\ref{psiEq}). Note that we can make translation invariance in Lagrangian space explicit by using $\q^{ab}=\q^{a}-\q^{b}=\q^{a1}-\q^{b1}$, and therefore $\q^1$ can be safely set to zero in $M$.  
In redshift space (\ref{M}) still holds after replacing $\sigma^2\mathdsOne\to\psi_{(r)}(\bm{0})$ and $\psi\to\psi_{(r)}$. This is valid for the plane-parallel case where $\hat r$ is a constant not depending on $\x^a$. For the angular case, one has to keep track towards which $\x_a$ the $\hat r$'s are pointing. We will make this dependence explicit only in the next section.

We can do the Gaussian integral above over $\tilde{\bm{K}}$ in (\ref{Gtemp}) to find
\be\label{Gprem}
G_n=\int \frac{d^{3n}\tilde Q}{(2\pi)^{3n/2}} \frac{1}{\sqrt{\det\left[M\right]}} e^{-\frac{1}{2}(\tilde{X}_i-\tilde{Q}_i) \left[M^{-1}(\tilde{\bm{Q}})\right]_{ij} (\tilde{X}_j-\tilde{Q}_j)}
\ee
Clearly, we could simply stop here and claim victory. However, the expression above is not explicitly translation invariant in Lagrangian space as it depends on $\tilde Q$. This results in the fact that for $n=2$, the expression above involves a 6-dimensional integral, while our previous expression for $\xi$, (\ref{xi}), involved just a 3-dimensional integral. So, let us do some more work and make translation invariance in Lagrangian space explicit. To do that we need to undo the integral in $\bm{K}$ and work with (\ref{Gtemp}).

\subsection{Making translation invariance in Lagrangian space explicit}

Let us subtract $\x^1$ from all $\x$'s,  and $\q^1$ from all $\q$'s, and split $\bm{K}$ in two pieces, and denote the resulting  vectors (of total length $3(n-1)$) as
\be\label{VecsRel}
\bm{X}^T&\equiv& \left((\x^{21})^T,(\x^{31})^T,\dots,(\x^{n1})^T\right)\nonumber\\
\bm{Q}^T&\equiv& \left((\q^{21})^T,(\q^{31})^T,\dots,(\q^{n1})^T\right)\nonumber\\
\bm{K}^T&\equiv& \left((\k^{2})^T,(\k^{3})^T,\dots,(\k^{n})^T\right)
\ee
Comparing with (\ref{Vecs}), one can  write:
\be\label{zeroPad}
\tilde{X}_j&=&\left(0,0,0,\bm{X}^T\right)_j+\left((\x^{1})^T,(\x^{1})^T,(\x^{1})^T,\dots\right)_j={X}_iW_{ij}+x^1_i T_{ij}\nonumber\\
\tilde{Q}_j&=&\left(0,0,0,\bm{Q}^T\right)_j+\left((\q^{1})^T,(\q^{1})^T,(\q^{1})^T,\dots\right)_j={Q}_iW_{ij}+q^1_i T_{ij}\nonumber\\
\tilde{K}_j&=&\left(0,0,0,\bm{K}^T\right)_j+\left((\k^{1})^T,0,0,0,\dots\right)_j={K}_iW_{ij}+k^1_i Z_{ij}
\ee
where by inspection we can read off $T$ to be a 3-by-$3n$ matrix defined by stacking $n$ (3-by-3) identity matrices ($\mathdsOne$):
\be\label{T}
T\equiv\left(\mathdsOne_{3\times 3},\mathdsOne_{3\times 3},\mathdsOne_{3\times 3},...,\mathdsOne_{3\times 3}\right)      
\ee
We defined $Z$ to be a 3-by-$3n$ matrix defined by stacking $1$ (3-by-3) identity matrix with (n-1) null matrices of size 3-by-3 :
\be\label{Z}
Z\equiv\left(\mathdsOne_{3\times 3},0_{3\times 3},0_{3\times 3},...,0_{3\times 3}\right)       
\ee
We also find  $W$ to be a $3(n-1)$-by-$3n$ matrix defined by
\be\label{W}
W\equiv\left(0_{3(n-1)\times 3},\mathdsOne_{3(n-1)\times 3(n-1)}\right)    
\ee

Making the substitutions given by (\ref{zeroPad}) in (\ref{Gtemp}), and changing integration variables according to $\tilde{\bm{Q}}\to(\q^1,\bm{Q})$ (the Jacobian of the transformation is 1),  we can do the integral in $\q^1$ by invoking  translation invariance in Lagrangian space, which tells us that $M$ is really a function of $\bm{Q}$ only, and not of $\tilde{\bm{Q}}$. The $\q^1$ integral gives a delta function setting $k^1_i=-{K}_kW_{kj}(T^T)_{ji}$. This looks obscure until we plug in the values of $W$ and $T$ to find $$k^1_i=-\sum_{a=2}^nk_i^a\ ,$$ which is nothing but the standard  ``the sum of $\k$'s should equal zero'' rule. 

Using this value of $\k^1$, one can check that $\x^1$, which goes in the exponent of (\ref{Gtemp}) through $\tilde{\bm X}$, drops out. Thus, in the angular case of redshift space, $\x^1$ appears only in the covariance $M$ through $\hat r$ in $\psi_{(r)}$. Remember that we obtained the same result for the 2-point function in redshift space as well.

Having eliminated the $\k^1$ and $\q^1$ integrals, we are left with a  Gaussian integral in $\bm{K}$. Performing it and using the expressions for $T$, $Z$ and $W$ one can show after a bit of linear algebra that $G_n$ is given by:
\be\label{nptFinal}
G_n(\x^1,\x^2,\dots)&=&\langle (1+\delta(\x^1))(1+\delta(\x^2))(1+\delta(\x^3))...(1+\delta(\x^n))\rangle\nonumber\\
&=&\int \frac{d^{3{(n-1)}} Q}{(2\pi)^{3(n-1)/2}} \frac{1}{\sqrt{\det\left[\mathcal{N}\right]}} e^{-\frac{1}{2}(X_i-Q_i) \left[\mathcal{N}^{-1}(\bm{Q})\right]_{ij} (X_j-Q_j)}
\ee
 $\bm{X}$ is  the $3(n-1)$-vector given by (\ref{VecsRel}),  which we copy here in its explicit form:
\be\label{VecsRelFinalX}
\bm{X}^T&=& \left((\x^2-\x^1)^T,(\x^3-\x^1)^T,...(\x^n-\x^1)^T\right)
\ee
Before we write down $\mathcal{N}$, let us redefine the $3(n-1)$-vector $\bm{Q}$ so as not to make a reference to the irrelevant $\q^1$:
\be\label{VecsRelFinalQ}
\bm{Q}^T&\equiv& \left((\bm{Q}^{1})^T,(\bm{Q}^{2})^T,...(\bm{Q}^{n-1})^T\right)
\ee
where each $\bm{Q}^a$ is a column vector of length 3. The $3(n-1)$-by-$3(n-1)$ covariance matrix $\mathcal{N}$ then consists of $(n-1)^2$ 3-by-3 blocks given by 
\be\label{NblockFinal}
\left(\mathcal{N}^{ab}\right)_{3\times 3}=\mathdsOne\sigma^2+\psi(\bm{Q}^{a}- \bm{Q}^{b})-\left(\psi\big(\bm{Q}^{a}\big)+ \psi\big(\bm{Q}^{b}\big)\right)
\ee
where $a$ and $b$ correspond to the block of $\mathcal{N}$ at position $(a,b)$. Thus, $a$ and $b$ go between 1 and $(n-1)$. 

In real space, we can safely set $\x^1=0$ above. In redshift space we recover (\ref{nptFinal}) but with the replacement $\mathcal{N}\to\mathcal{N}_{(r)}$, where as one can easily guess:
\be\label{NblockFinalRS}
\left(\mathcal{N}_{(r)}^{ab}\right)_{3\times 3}&=&\psi_{(r)}(\bm{0},\x^1,\x^1)+\psi_{(r)}(\bm{Q}^{a}- \bm{Q}^{b},\x^{a+1},\x^{b+1})-\nonumber\\
&&-\left(\psi_{(r)}\big(\bm{Q}^{a},\x^{a+1},\x^1\big)+ \psi_{(r)}\big(\bm{Q}^{b},\x^1,\x^{b+1}\big)\right)
\ee

Let us check that we recover the result for the 2-point function. For $n=2$, $\mathcal{N}$ and $\mathcal{N}_{(r)}$ are both 3-by-3 matrices, and so they have only one 3-by-3 block each, with $a=b=1$ in (\ref{NblockFinal}) and (\ref{NblockFinalRS}), respectively. Clearly then, one obtains $\mathcal{N}=N$ and $\mathcal{N}_{(r)}=N_{(r)}$, and therefore (\ref{nptFinal}) reproduces (\ref{xi}). 

\subsection{Summary}

To tersely summarize, our results are given by (\ref{nptFinal}) with the definitions (\ref{VecsRelFinalX}) and (\ref{VecsRelFinalQ}). For real space one should use the real-space displacement covariance, given by  (\ref{NblockFinal}), while for redshift space one should instead substitute $\mathcal{N}\to\mathcal{N}_{(r)}$  in (\ref{nptFinal}) and use (\ref{NblockFinalRS}). The results are valid for both the plane-parallel as well as for the angular case in redshift space, with the line-of-sight vector $\hat r$ entering through $\psi_{(r)}$ defined through (\ref{psir}). The fact that translation invariance is broken in the angular case is captured completely by $\hat r$ entering through $\psi_{(r)}$.

Let us compare our final result, (\ref{nptFinal}), with our  initial much more easily derived result, (\ref{Gprem}).  We see that (\ref{nptFinal}) explicitly shows translation invariance in Lagrangian space and involves an integral with dimensions smaller by 3. Reducing the dimensionality of the integral by that much results in  an enormous speed-up when evaluating it numerically, thus justifying the algebra of the previous section.

\section{Numerical implementation}\label{sec:numdisc}

In the previous section we gave the explicit expression for the moments $\langle (1+\delta)^n\rangle$ -- equation (\ref{nptFinal}). The integral in (\ref{nptFinal}) is over a positive-definite function and so can be done easily numerically. However, extracting the connected $n$-point functions  $\langle \delta^n\rangle_c$ knowing the moments $\langle (1+\delta)^m\rangle$ (with $m\leq n$), albeit trivial to write down, involves subtracting comparable quantities all of which are $\mathcal{O}(1)$ to obtain a quantity which is $\ll 1$. 

So, for the 2-point function, for example, (\ref{xi}) gives $1+\xi$ as an integral over a positive definite function. However, to get $\xi$ one needs to subtract $1$ from that integral, which is a recipe for a numerical disaster when $\xi$ is small and comparable to round-off errors -- e.g. at large separations or high redshift. 

This problem is  especially exacerbated when one goes to higher and higher $n$-points functions outside the non-linear regime, when applying the ZA makes sense. We can trace the problem to (\ref{beginning}), which tells us that unlike Eulerian perturbation theory, Lagrangian perturbation theory gives us the \textit{total} density, from which we need to subtract the mean to get the overdensity. 

This paper does not solve this problem. Instead it ameliorates it by using the following trick. Let us first focus on $\xi$ in (\ref{xi}). The $1$ (i.e. the disconnected piece) can be plugged back into the $q$ integral by rewriting it as the right hand side of  (\ref{xi}) but with an $N$ given by $2\sigma^2$ times the identity matrix. So, the disconnected piece is given by the rhs of (\ref{xi}) after setting the cross term, $\psi=\langle \s(\bm{0})\s(\bm{q})\rangle_c$, to zero, thus making the two points ($\bm{0}$ and $\bm{q}$) uncorrelated, i.e. disconnected. This comes at a cost. The integrand in (\ref{xi}) will then no longer be positive definite. However, at large $q$, when $\psi\to0$, the integrand will quickly fall to zero. 

The same trick can be easily applied for extracting higher order $n$-point functions. For each disconnected piece, one has to use the same integrand as the one in (\ref{nptFinal}) but after setting the  $\psi$'s between the  disconnected points to zero.

The above trick is implemented in ZelCa both for the 2-point and for the 3-point functions. ZelCa uses the Cuba\footnote{\url{http://www.feynarts.de/cuba/}} \cite{2005CoPhC.168...78H} library for multidimensional numerical integration, as well as the Eigen\footnote{\url{http://eigen.tuxfamily.org}} C++ template library for linear algebra, which makes the ZA part of the code especially readable and easy to understand and modify as needed. We also provide a Sage\footnote{\url{http://www.sagemath.org/}} code for displaying the results, and also for calculating the tree-level 3-point function used in the next section.

\section{Numerical results}\label{sec:nums}

\subsection{The 2-dimensional redshift-space 2-pt correlation function}

Results from ZelCa for the real-space 2-point function, as well as for the redshift-space monopole, quadrupole and hexadecapole, were already included in Figure 3 of  \cite{2013arXiv1311.4884T}. By comparing them to the N-body results from \cite{2013MNRAS.429.1674C}, there we showed that the ZA and the N-body results show excellent agreement at large scales ($\gtrsim 20$Mpc$/h$) relevant for the BAO (see \cite{2013arXiv1311.4884T} for further details).

\begin{figure}[h!]
\centering
    \subfloat{\includegraphics[width=0.49\textwidth]{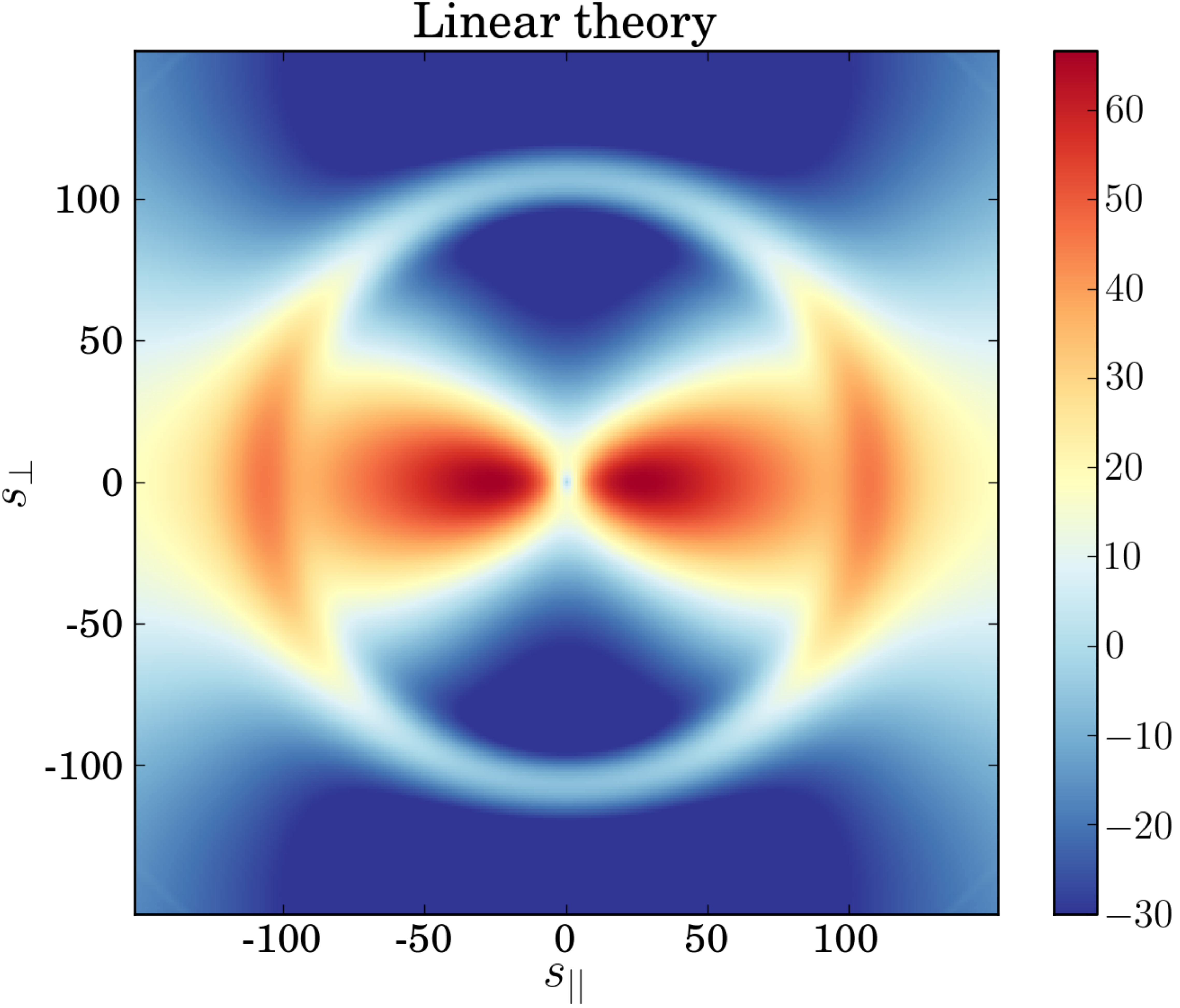}}\hfill
  \subfloat{\includegraphics[width=0.49\textwidth]{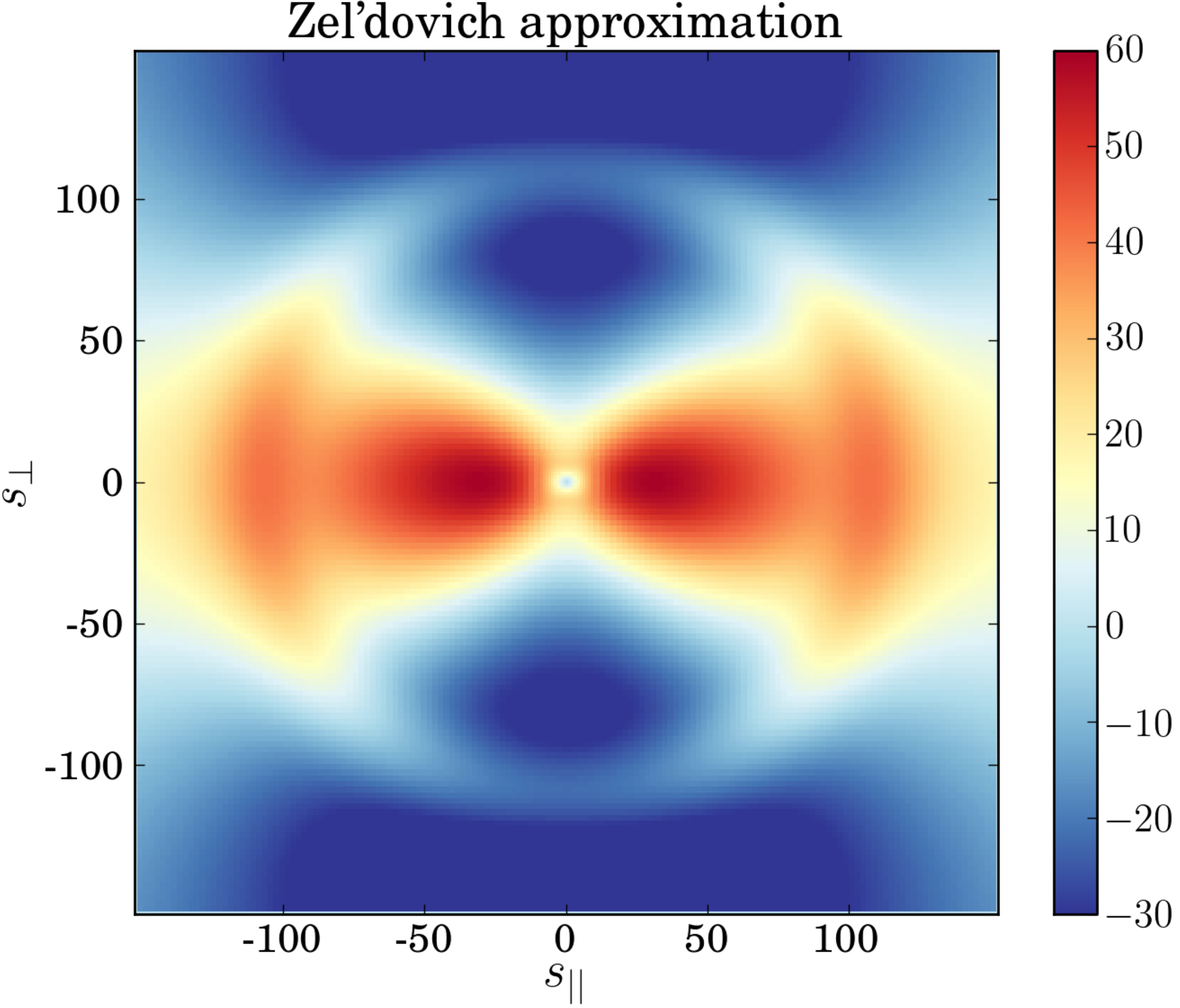}}
\caption{The 2-dimensional 2-point correlation function (at $z=0.35$, chosen to match \cite{2012MNRAS.427.2132P}) in redshift space, as predicted by linear SPT and by the ZA. The correlation function is scaled by $s_{||}^2+s_{\bot}^2$.  One should compare these results with the same plot obtained from mock catalogs: upper-right panel of Figure~4 in \cite{2012MNRAS.427.2132P}. Our results do not include bias, and therefore quantitative differences remain. However, qualitatively, one can see that linear theory predicts features which are too sharp, while the ZA which captures the smearing effects of the bulk flows, results in smooth features well in line with the realistic mock catalog results. Note that unlike the main text, $s$ here denotes separations (not displacements) along ($||$) and perpendicular ($\bot$) to the line of sight -- a notation chosen to match that of \cite{2012MNRAS.427.2132P}.} \label{fig:rs}
\end{figure}

In this paper, we show the first results for the 2-dimensional redshift-space 2-point correlation function -- the right panel of Figure~\ref{fig:rs}. In the left panel of that figure we show the result of linear Standard Perturbation Theory (SPT) (e.g. \cite{2002PhR...367....1B}). The most notable difference between the two is the fact that the ZA produces a relatively smoother 2-pt function -- a direct consequence of the smearing effects of the bulk flows, which the ZA captures, while linear theory does not.

One can also compare the figure to the upper-right panel of Figure~4 in \cite{2012MNRAS.427.2132P}, which shows results obtained using the LasDamas (Large suit of Dark matter simulations) mock catalogs\footnote{\url{http://lss.phy.vanderbilt.edu/lasdamas/}}. We do not include Lagrangian bias in this calculation, and therefore a detailed quantitative comparison is not possible. Yet, even by eye one can see that the  features in the mock catalog calculation are smoother than the linear theory result, and on par with the ZA prediction.


This should not come as a surprise -- as already discussed at length in the companion to this paper, \cite{2013arXiv1311.4884T}, we already expected that at the  scales of the BAO, linear theory should receive $\mathcal{O}(1)$ corrections in real space, while the ZA should be correct to within $\sim 3\%$ for the matter density (again in real space). The discussion in Section~4.2 of that paper, shows that the ZA  performs extremely well compared to linear theory for redshift space as well.  Thus, Figure~1 serves as yet another confirmation of those results.

\subsection{The real space 3-pt function}

\begin{figure}[h!]
\centering
\includegraphics[width=0.9\textwidth]{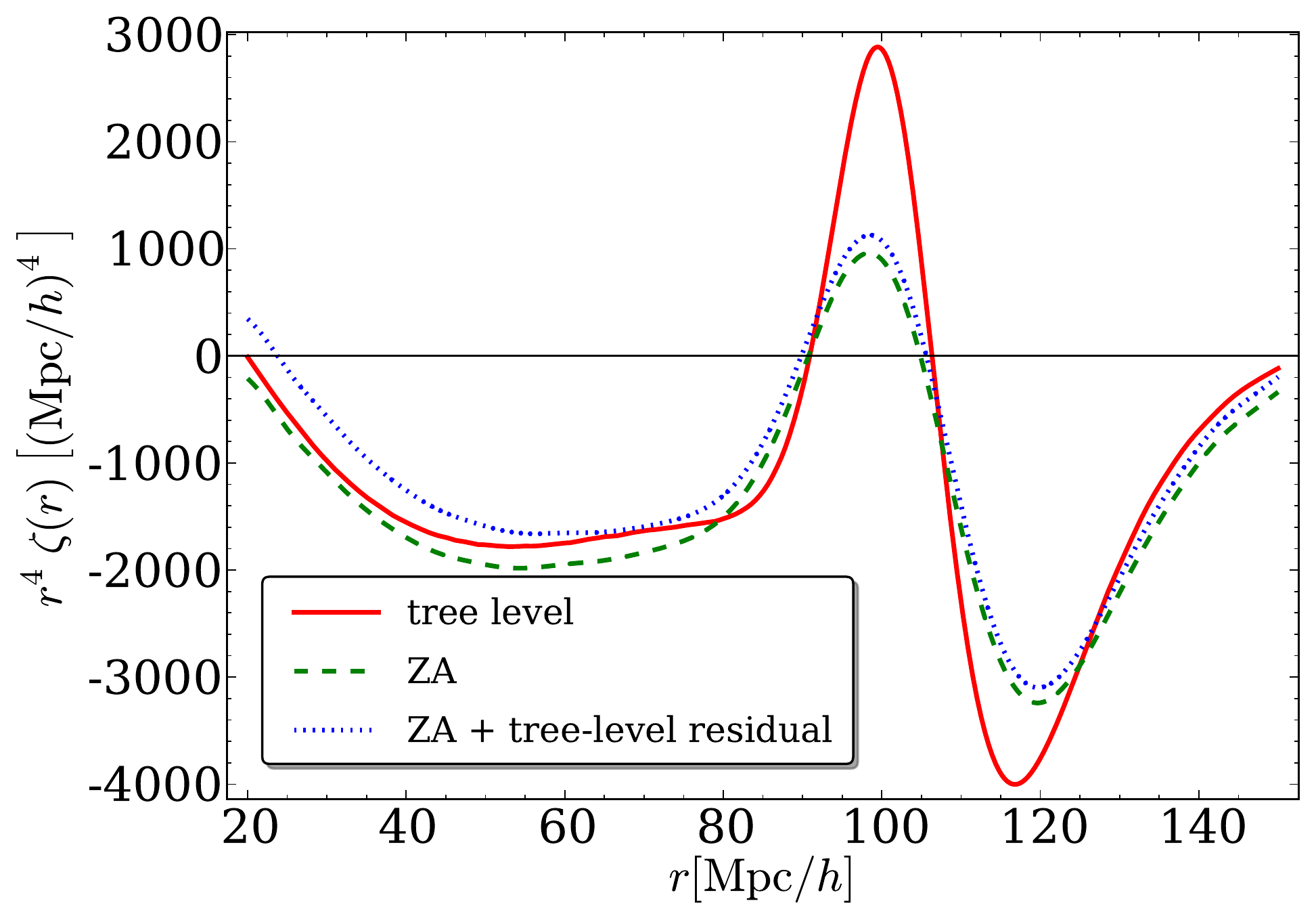}
\caption{The real-space 3-point correlation function (at $z=0.55$, chosen to match  \cite{2013arXiv1311.4884T} and \cite{2013MNRAS.429.1674C}) for equilateral triangles, as predicted by the ZA and at tree-level by SPT. In the ZA, one can clearly see the smearing in the acoustic peak due to the bulk flows, similar to what one finds for the 2-pt function. We also show the ZA result, corrected for the missing tree-level contributions. One should note the small difference between the ZA and the corrected ZA results. See the text for further discussion.} \label{fig:3pt}
\end{figure}

Even though we only focus on the 2-pt function in \cite{2013arXiv1311.4884T}, it is clear that a similar analysis should apply to higher-order statistics -- with the ZA capturing the main effects of the large-scale coherent motions in the universe. 
To check that intuition, in Figure~\ref{fig:3pt} we show results for the real-space 3-pt function, $\zeta$, in the equilateral configuration. One can clearly see the effects of the bulk flows smearing the BAO peak when compared to the tree-level prediction (as given in e.g. \cite{2002PhR...367....1B}).

One can immediately object, however, that the ZA does not capture the tree-level result for $\zeta$, given that the overdensity it predicts matches SPT only at the linear level. Therefore, we also calculate the following quantity:
\be
\tilde \zeta_{\mathrm{ZA}}\equiv \zeta_{\mathrm{ZA}} + \left.\bigg(\zeta_{\mathrm{SPT}}-\zeta_{\mathrm{ZA}}\bigg)\right|_{\mathrm{at\ tree-level}}
\ee
which includes the piece of the tree-level, which the ZA misses. We plot $\tilde \zeta_{\mathrm{ZA}}$ in Figure~\ref{fig:3pt} as well (line denoted ``ZA + tree-level residual''). It is quite interesting to note how close $\tilde \zeta_{\mathrm{ZA}}$ is to $\zeta_{\mathrm{ZA}}$ (within $\sim$10\%), i.e. how small the piece of the tree-level result that the ZA misses is.\footnote{The tree-level prediction for $\zeta$ in SPT is given by eq.(157) of \cite{2002PhR...367....1B}. It is obtained using the standard perturbation theory $F_2$ kernel, given in e.g. \cite{1996ApJS..105...37S}. Note that \cite{1996ApJS..105...37S} also gives the $F_2$ kernel in the ZA, $F_2^{\mathrm{ZA}}$. Comparing the two, we see that the ZA captures the coefficients of the four terms of the tree-level $\zeta$, appearing in eq.(157) of \cite{2002PhR...367....1B}, with  errors of  30\%, 0, 0 and 75\%, respectively.} At the same time, the difference between the tree-level and ZA results is at $\mathcal{O}(1)$ around the BAO peak, in direct analogy with the 2-pt correlation function results \cite{2013arXiv1311.4884T}.

The  difference between $\zeta_{\mathrm{ZA}}$ and $\tilde \zeta_{\mathrm{ZA}}$ can be thought of as the typical error one makes by using the ZA. That difference being small, in what follows, we will assume that $\tilde \zeta_{\mathrm{ZA}}$ is close to the true result. Whether that is indeed the case remains to be seen, but we consider it our best guess in light of \cite{2013arXiv1311.4884T}. 

Then the fact that $\zeta_{\mathrm{ZA}}$ is so close to the truth (as approximated by $\tilde \zeta_{\mathrm{ZA}}$) can be qualitatively understood by remembering that CDM particle trajectories are extremely well-captured by the ZA (see \cite{2013arXiv1311.4884T}, and in particular their Section 8). Therefore, the predicted density field using the full unexpanded ZA is quite close to the true density field (e.g. Fig. 1 in \cite{2013arXiv1311.4884T}). Indeed, one finds that the cross-correlation coefficient between the true  and  ZA-predicted density fields is close to 1 well into the non-linear regime, unlike the SPT prediction, which decorrelates from the truth at relatively large scales  \cite{2012JCAP...04..013T,1993MNRAS.260..765C}. A good cross-correlation coefficient has already been qualitatively shown to imply  well-recovered $n$-point functions \cite{2013JCAP...06..036T,2013arXiv1311.4884T}. So, it is not surprising that if one keeps the ZA unexpanded (as emphasied in \cite{2013arXiv1311.4884T}), one recovers a very good approximation for the density 3-pt statistics at large scales.

\section{Summary}\label{sec:summary}

In this paper we presented an expression for the matter $n$-point functions in the Zel'dovich approximation in both real and redshift space, with the results for the latter valid both for   the plane-parallel limit, as well as for the angular case. We also discussed the numerical implementation of our results and provide the code, called ZelCa, generating the results of this paper. 

Using that code we obtained numerical results for the 2-dimensional redshift-space 2-pt correlation function. We find a qualitative match (as we do not include bias) between the ZA prediction and the result obtained from mock catalogs, while the tree-level result gives a BAO feature which is too sharp, as already expected from the results for the redshift-space multipoles \cite{2013arXiv1311.4884T}. 

We also show results for the equilateral configuration for the 3-pt function around the acoustic peak and demonstrate the expected smearing of the BAO due to bulk motions, similar to the smearing observed for the 2-pt function. We  calculate the tree-level correction to the ZA result and show it to contribute only at $\mathcal{O}(0.1)$, reassuringly in line with our analysis in \cite{2013arXiv1311.4884T}.

The road from here on involves adding more realism to the ZA. Fortunately, a lot of progress in that direction was already done for the 2-pt function in \cite{2013MNRAS.429.1674C}. Those authors showed how one can include Lagrangian bias and higher order corrections to the 2-pt function in the ZA. Their analysis can be transparently followed for the higher $n$-point functions as well. However, one should always keep in mind that short scales may introduce non-negligible corrections to the analytical perturbative results, which need to be calibrated from simulations through the effective field theory formalism (see \cite{2013arXiv1311.2168P}).

\bibliography{mildly_NL_v2}

\end{document}